\documentclass[twocolumn,english,reprint,superscriptaddress,prl]{revtex4-1}

\usepackage[T1]{fontenc}
\usepackage[latin9]{inputenc}
\usepackage{color}
\usepackage{bm}
\usepackage{amstext}
\usepackage{amssymb}
\usepackage{graphicx}
\usepackage{booktabs}
\usepackage{tabularx}
\usepackage{amsmath,mathrsfs}
\makeatletter

\DeclareMathOperator*{\argmin}{argmin}   

\newcommand{\re}{\mathbb{R}\text{e}}
\newcommand{\im}{\mathbb{I}\text{m}}
\usepackage{babel}

\usepackage{hyperref}
\hypersetup{
 linktocpage = true,
    colorlinks,
    citecolor=blue,
    filecolor=black,
    linkcolor=blue,
    urlcolor=blue
}

\usepackage{wasysym}

\makeatother

\begin{document}
\author{Giacomo Torlai}
\affiliation{Center for Computational Quantum Physics, Flatiron Institute, New York, New York, 10010, USA}

\author{Juan Carrasquilla}
\affiliation{Vector Institute for Artificial Intelligence, MaRS Centre, Toronto, ON, Canada M5G 1M1}
\affiliation{Department of Physics and Astronomy, University of Waterloo, Ontario, N2L 3G1, Canada}

\author{Matthew T. Fishman}
\affiliation{Center for Computational Quantum Physics, Flatiron Institute, New York, New York, 10010, USA}

\author{Roger G. Melko}  
\affiliation{Department of Physics and Astronomy, University of Waterloo, Ontario, N2L 3G1, Canada}
\affiliation{Perimeter Institute for Theoretical Physics, Waterloo, Ontario N2L 2Y5, Canada}

\author{Matthew P. A. Fisher}
\affiliation{Department of Physics, University of California, Santa Barbara, CA 93106, USA}

\title{Wavefunction Positivization via Automatic Differentiation}

\begin{abstract}
We introduce a procedure to systematically search for a local unitary transformation that maps a wavefunction with a non-trivial sign structure into a positive-real form. The transformation is parametrized as a quantum circuit compiled into a set of one and two qubit gates. We design a cost function that maximizes the average sign of the output state and removes its complex phases. The optimization of the gates is performed through automatic differentiation algorithms, widely used in the machine learning community. We provide numerical evidence for significant improvements in the average sign for a two-leg triangular Heisenberg ladder with next-to-nearest neighbour and ring-exchange interactions. This model exhibits phases where the sign structure can be removed by simple local one-qubit unitaries, but also an exotic Bose-metal phase whose sign structure induces ``Bose surfaces'' with a fermionic character and a higher entanglement that requires deeper circuits.
\end{abstract}

\maketitle

\selectlanguage{english}
\paragraph{Introduction.} 

The most striking contrast between the classical and quantum world is the fact that quantum wavefunctions contain ``probability'' amplitudes that are not strictly real and positive.
This so-called {\it sign} (or phase) structure is an essential feature of a variety of quantum phenomena with no classical counterpart, such as the Pauli exclusion principle, entanglement, and quantum interference. It lies at the heart of any algorithm for quantum computing~\cite{benenti}.

A sign structure often hinders the simulation of quantum many-body states by means of classical resources, and it essentially defines the 
threshold for what can be considered truly quantum-mechanical. Indeed, there is a one-to-one mapping between a real, {\it non-negative} wavefunction and a classical probability distribution, formulated explicitly by the Born rule. However, the sign structure is not a universal feature of a quantum state, since it strongly depends on the choice of basis. As such, for a given state it is only natural to wonder: is there a {\it local} change of basis that removes the sign structure, leading to a non-negative wavefunction?

Given a preferred ``computational basis'', finding and applying a change of basis involves implementing a unitary transformation.  
The resources required for this task can however be non-trivial.
For example, any ground state becomes non-negative in the energy eigen-basis, 
but finding the corresponding (non-local) unitary transformation is equivalent to diagonalization, with a complexity that scales exponentially in the number of qubits. 
The question becomes, can a change of basis be discovered with a transformation represented by a local unitary circuit of small depth?

 Such transformations are typically identified based on simple physical principles related to the structure of the Hamiltonian and its symmetries. 
The most notable example is the Marshall sign rule~\cite{Marshall}, eliminating the sign structure from the ground states of quantum antiferromagnets on bipartite lattices. 
The resulting theoretical insight means that new bases that simplify the sign structure for a specific frustrated magnet or fermion model are routinely discovered~\cite{MajBasis,Kaul13,Wessel17,Honecker16}. In turn, in a few instances it has also been rigorously proven that efficient transformations do not exist~\cite{Hastings16,Ringel}, rendering the sign structure ``intrinsic''. However, if no obvious transformation is known, it is generally unclear whether the offending sign structure is intrinsic or whether it only persists due to a lack of physical insight.
An automated procedure to search for relevant transformations is therefore highly desirable.

In this paper, we propose an algorithm to tackle this question which combines tensor networks and differentiable programming. We formulate the search for the local basis as an optimization task over quantum circuits compiled into a set of local quantum gates. By optimizing a suitable cost function, a quantum circuit is used to positivize a quantum state with a sign structure. We show how this procedure can be realized in practice by adopting a tensor network representation of the quantum circuit, and applying automatic differentiation to obtain a ``learning signal'' for each quantum gate. We present a proof-of-principle demonstration for a two-leg triangular Heisenberg ladder with four-spin ring exchange interaction, which harbors a sign structure of tunable complexity,
including that of an exotic highly entangled spin Bose-metal phase.

\paragraph{Learning a sign structure.} We study a system composed of $N$ qubits described by a wavefunction $|\Psi\rangle$. For a given choice of basis of the many-body Hilbert space $|\bm{\sigma}\rangle=|\sigma_1,\dots,\sigma_N\rangle$, we assume that the wavefunction has a sign structure, i.e.~the coefficients $\Psi(\bm{\sigma})=\langle\bm{\sigma}|\Psi\rangle$ appear with both positive and negative signs. We note that, while we restrict to real wavefunctions, the following approach identically applies to the case where the wavefunction is complex-valued.

\begin{figure}[t]
\noindent \centering{}\includegraphics[width=0.8\columnwidth]{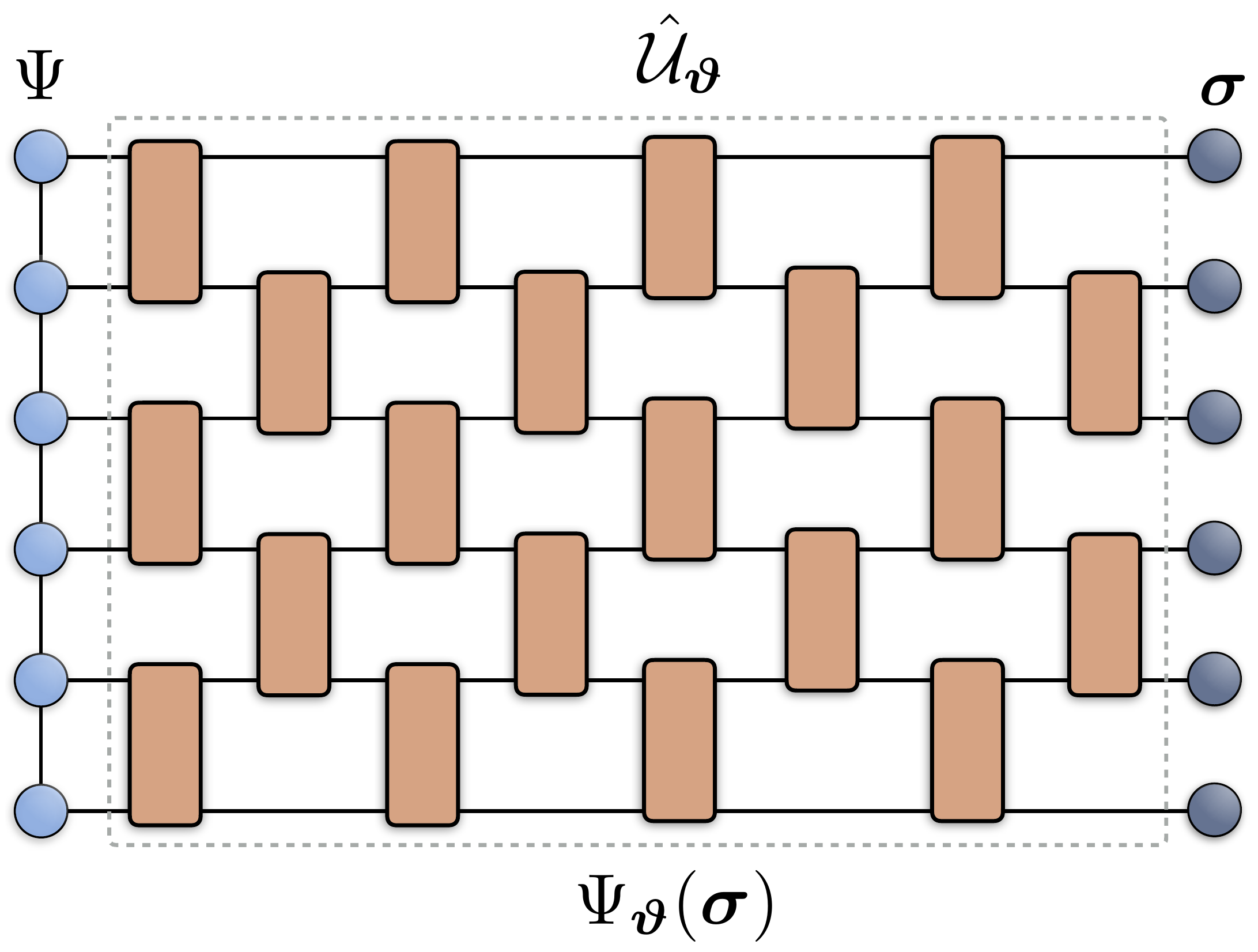}
\caption{({\bf a}) Projection of the wavefunction on the basis state $\langle\bm{\sigma}|$ after the application of the quantum circuit $\hat{\mathcal{U}}(\bm{\vartheta})$ implementing the change of basis. Here, the circuit is compiled into a set of local two-qubit gates.}
\label{Fig::qc} 
\end{figure}

Given the sign structure $\text{Sign}\big(\Psi(\bm{\sigma})\big)$, how can we run an automated search for a local unitary transformation $\hat{\mathcal{U}}$ generating a non-negative wavefunction? For this purpose, it is natural to express the unitary as a quantum circuit, where locality can be imposed at the level of the quantum gates (Fig.~\ref{Fig::qc}). Because of their universality~\cite{benenti}, we can restrict to single- and two-qubit gates acting on pairs of adjacent sites. Then, the unitary transformation is written in terms of parameters $\bm{\vartheta}=\{\bm{\vartheta}^{[1]},\bm{\vartheta}^{[2]},\dots\}$, where $\bm{\vartheta}^{[k]}$ are a set of real and continuous parameters characterizing each single gate.

Starting from a wavefunction $\Psi(\bm{\sigma})$ displaying a sign structure, provided as input to the quantum circuit, the goal is to discover a set of gates such that the output state is non-negative. We choose to phrase this problem as an optimization task, where the non-negativity of the output state is enforced upon minimizing a suitable {\it cost function} $\mathcal{C}({{\bm{\vartheta}}})$. More precisely, the optimal set of parameters $\bm{\vartheta}^*=\argmin_{\bm{\vartheta}}\mathcal{C}({{\bm{\vartheta}}})$ should satisfy $\Psi_{\bm{\vartheta}^*}(\bm{\sigma})\ge0\:\:\forall|\bm{\sigma}\rangle$, where $\Psi_{\bm{\vartheta}^*}(\bm{\sigma})=\langle\bm{\sigma}|\,\hat{\mathcal{U}}_{\bm{\vartheta}^*}|\Psi\rangle$. Given some initial configuration of the circuit, the optimization is solved by iteratively updating the gates according to the gradient of the cost function, $\bm{\vartheta}\rightarrow\bm{\vartheta}-\eta\,\mathcal{G}({{\bm{\vartheta}}})$, where $\mathcal{G}({{\bm{\vartheta}}})=\nabla_{\bm{\vartheta}}\mathcal{C}({{\bm{\vartheta}}})$  and $\eta$ is the step-size of the update (often called {\it learning rate}). More sophisticated algorithms developed within the machine learning community can also be implemented, such as the adaptive learning rates~\cite{adadelta,adam} or higher-order gradients~\cite{Amari97}.

The cost function is the most crucial ingredient. On one hand, it needs to correctly capture the objective of the optimization. On the other hand, the sign structure is a global property of the quantum state, and thus the calculation of the cost function (and its gradients) should also remain scalable with the number of qubits. For the latter, it is prudent to express $\mathcal{C}({{\bm{\vartheta}}})$ as an expectation value over the probability distribution underlying the quantum state at the output of the circuit:
\begin{equation}
\mathcal{C}({\bm{\vartheta}})=\sum_{\bm{\sigma}}|\Psi_{\bm{\vartheta}}(\bm{\sigma})|^2\mathcal{C}_{{\bm{\vartheta}}}(\bm{\sigma})\:.
\label{Eq::cost}
\end{equation}
In fact, provided one can sample the distribution \linebreak $p_{\bm{\vartheta}}(\bm{\sigma})=|\Psi_{\bm{\vartheta}}(\bm{\sigma})|^2$, the expectation value of Eq.~\eqref{Eq::cost} can be approximated with a sum over a finite number of configurations $\{\bm{\sigma}_j\}$ drawn from  $p_{\bm{\vartheta}}(\bm{\sigma})$. Now, the only task that remains is designing an appropriate function $\mathcal{C}_{{\bm{\vartheta}}}(\bm{\sigma})$.

Besides the sign of the wavefunction, an additional constraint that should be taken into account is that the complex phases, necessarily accumulated by a universal gate set, are eliminated by the end of the unitary evolution.
To capture both conditions on the imaginary part and the sign, it is convenient to split the cost function into a convex sum of two contributions:
\begin{equation}
\mathcal{C}_{{\bm{\vartheta}}}(\bm{\sigma})=\gamma\big|\im\big(\Psi_{\bm{\vartheta}}(\bm{\sigma})\big)\big|-
(1-\gamma)\text{Sign}(\mathbb{R}\text{e}(\Psi_{\bm{\vartheta}}(\bm{\sigma}))\:,
\end{equation}
where $\gamma\in[0,1]$. By tuning the parameters according to the gradient $\mathcal{G}({{\bm{\vartheta}}})$, the quantum circuit will try to increase the sign of the real part of the wavefunction, while forcing the imaginary part to be zero. Note that the initial average sign can always be set to a positive value by an appropriate global transformation.

\paragraph{Differentiable programming.} 
Next, in order to evaluate the gradients of the cost function we need to adopt a representation of the input quantum state and the quantum circuit amenable to scalable simulations. To this end, we assume that the initial state admits an efficient matrix product state (MPS) representation, and obtain the final state by contracting the MPS with the various gates in the circuit. At each intermediate step, provided the circuit depth is not too large, the quantum state can be restored into an MPS form by means of singular value decompositions.

The calculation of the gradients is the most involved step in the procedure, and analytical approaches would clearly be intractable. We leverage
{\it automatic differentiation} (AD) techniques~\cite{AD2000}, routinely used in machine learning applications to train neural-network architectures~\cite{Hinton_BP} and recently applied to optimize tensor network states~\cite{Liao19}. The core object in AD is the computational graph implementing the set of elementary computations (edges) acting on the variables (nodes). We specifically implement {\it reverse-accumulation} AD, where a forward pass first calculates the output of the graph, and derivates are calculated starting from the output, and back-propagated through the graph using a sequence of Jacobian-vector products. 

\begin{figure}[t]
\noindent \centering{}\includegraphics[width=1\columnwidth]{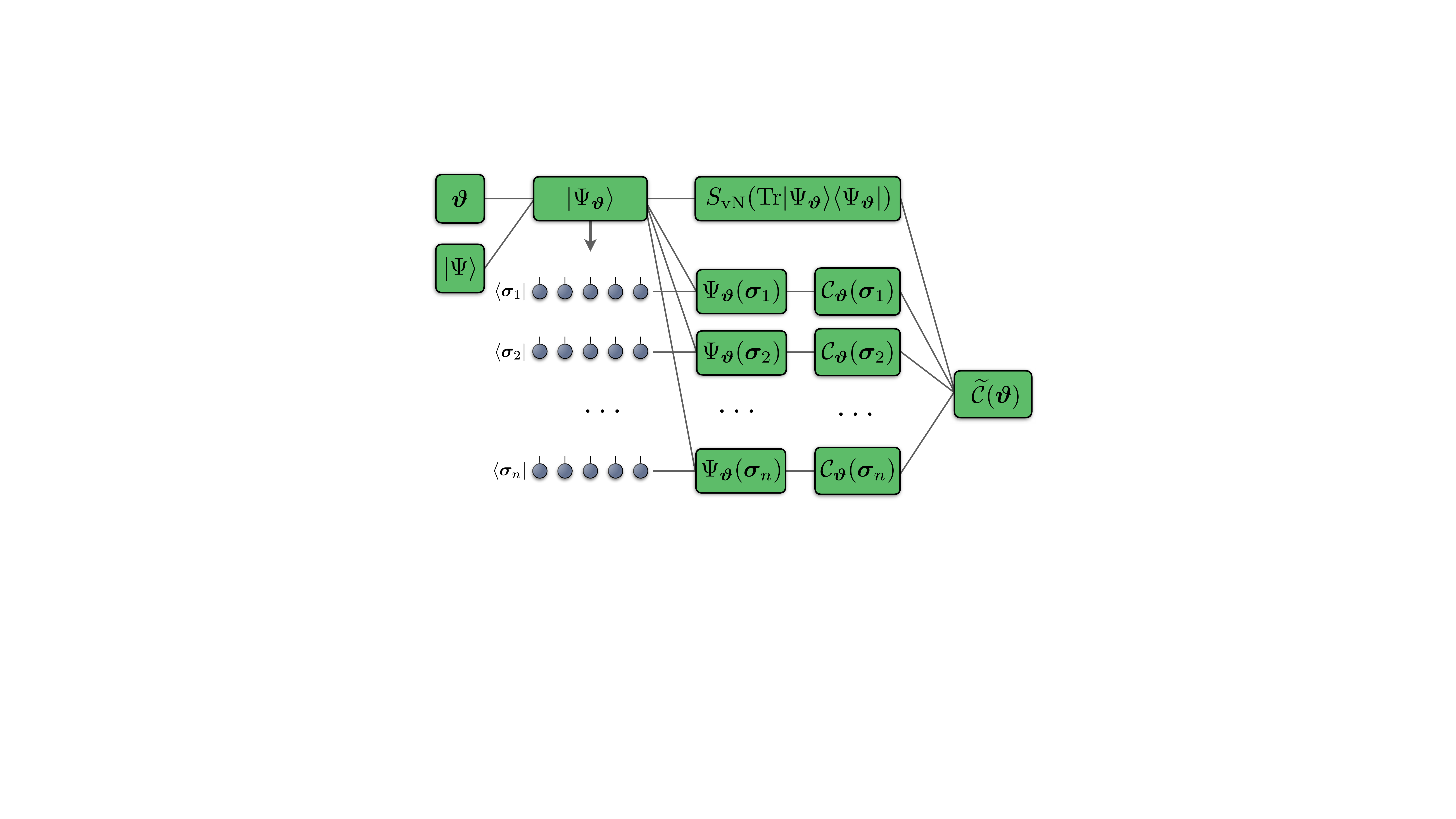}
\caption{
Schematic of the computational graph for the calculation of the cost function. The output quantum state $|\Psi_{\bm{\vartheta}}\rangle$, obtained by contracting the circuit tensor network, is sampled to generate the configurations $\{\bm{\sigma}_j\}$, which are used to compute the cost function. In addition, the entanglement entropy of the output state is added as a regularization to mitigate the growth of entanglement in deep circuits.}
\label{Fig::AD} 
\end{figure}

The computational graph implementing the positivization is divided into three stages (Fig.~\ref{Fig::AD}). First, the circuit $\hat{\mathcal{U}}_{\bm{\vartheta}}$ is applied to the input state through a series of tensor contractions. The resulting output quantum state $|\Psi_{\bm{\vartheta}}\rangle=\hat{\mathcal{U}}_{\bm{\vartheta}}|\Psi\rangle$ is then sampled to generate a set of $n$ configurations $\{\bm{\sigma}_j\}$ approximating the sum in Eq.~(\ref{Eq::cost}). The projections of $|\Psi_{\bm{\vartheta}}\rangle$ into these configurations are used to estimate the sample-wise cost function
\begin{equation}
\widetilde{\mathcal{C}}(\bm{\vartheta})=\frac{1}{n}\sum_{j=1}^n\mathcal{C}_{\bm{\vartheta}}(\bm{\sigma}_j)+\alpha S_{\text{vN}}(\hat{\rho}_{A})\:.
\end{equation}
Note that we have also added a term proportional to the entanglement entropy $S_{\text{vN}}(\hat{\rho}_A)=-\text{Tr}(\hat{\rho}_A\log\hat{\rho}_A)$, where $\hat{\rho}_{A}$ is the reduced density matrix for a equal bipartition of the qubits and $\alpha$ is a small weight. We introduce this type of regularization to the cost function to limit the growth of entanglement generated by the application of the gates, particularly relevant in the optimization of deep quantum circuits. Once the computational graph is compiled, the reverse-accumulation step evaluates the derivatives with respect to each gate parameter in the circuit (see Supplementary Material for more details).

\paragraph{Results.} We focus on the ground state wavefunctions of a two-leg triangular ladder with  Hamiltonian
\begin{equation}
\begin{split}
\hat{H} &= J_1\sum_j\bm{\hat{S}}_j\cdot\bm{\hat{S}}_{j+1}+J_2\sum_{j}\bm{\hat{S}}_j\cdot\bm{\hat{S}}_{j+2}\\
&+\frac{J_r}{2}\sum_j\bm{\hat{P}}_{j,j+1,j+3,j+2}+\bm{\hat{P}}^\dagger_{j,j+1,j+3,j+2}\:,
\end{split}
\label{BM}
\end{equation}
where $\bm{\hat{S}}_j$ are spin-$1/2$  operators.  Here, the ring-exchange term corresponds to the cyclic exchange of spin states, $\bm{\hat{P}}_{i,j,k,l}|S^z_i,S^z_j,S^z_k,S^z_l\rangle=|S^z_l,S^z_i,S^z_j,S^z_k\rangle$ and the couplings are $J_1=1$, $J_2,J_r>0$. 
The model in Eq.~\eqref{BM} exhibits a range of ground states with a sign structure of tunable complexity, so it serves as a representative testbed for our experiments. Whereas for $J_1=J_r=0$ or $J_2=J_{r}=0$ the sign of the ground state wavefunction can be eliminated via a unitary transformation acting on single spins~\cite{capriotti2001}, for $J_{r}/J_1\gg1$ and  $J_2/J_1\ll 1$ the model displays an exotic spin Bose-Metal (SBM) phase endowed with a complex sign structure associated with the presence of singular wave vectors or ``Bose surfaces''~\cite{Sheng09}. After obtaining the ground state MPS using standard density matrix renormalization group techniques~\cite{white_dmrg,ITensor},
we implement the AD graph using the machine learning library TensorFlow~\cite{tensorflow}. 

We first consider the case of $J_r=0$, corresponding to the one-dimensional $J_1$-$J_2$ model. In the limit of $J_2=0$ we recover the Heisenberg model, where the sign structure of the ground state $\Psi(\bm{S}^z)$ in the Ising basis $|\bm{S}^z\rangle=|S_1^z,\dots,S_N^z\rangle$ follows the Marshall sign rule~\cite{Marshall,capriotti2001}. The transformation removing this sign structure can be composed as a set of $N/2$ rotations of angle $\pi$ about the $z$ axis, corresponding to a depth-one quantum circuit. To check if this can be recovered by our procedure, we construct the variational quantum circuit $\hat{\mathcal{U}}_{\bm{\vartheta}}$ using one layer of single-qubit rotations around the $z$ axis. We run the positivization procedure for a chain containing $N=40$ spins. After randomly initializing the circuit parameters (i.e.~$N=40$ angles) we train the circuit to minimize the cost function using $n_S=10^3$ configurations sampled from the final MPS distribution $|\Psi_{\bm{\vartheta}}(\bm{S}^z)|^2$~\cite{White_mps,Ferris}, and update the parameters using the Adam optimizer~\cite{adam}.

\begin{figure}[t]
\noindent \centering{}\includegraphics[width=\columnwidth]{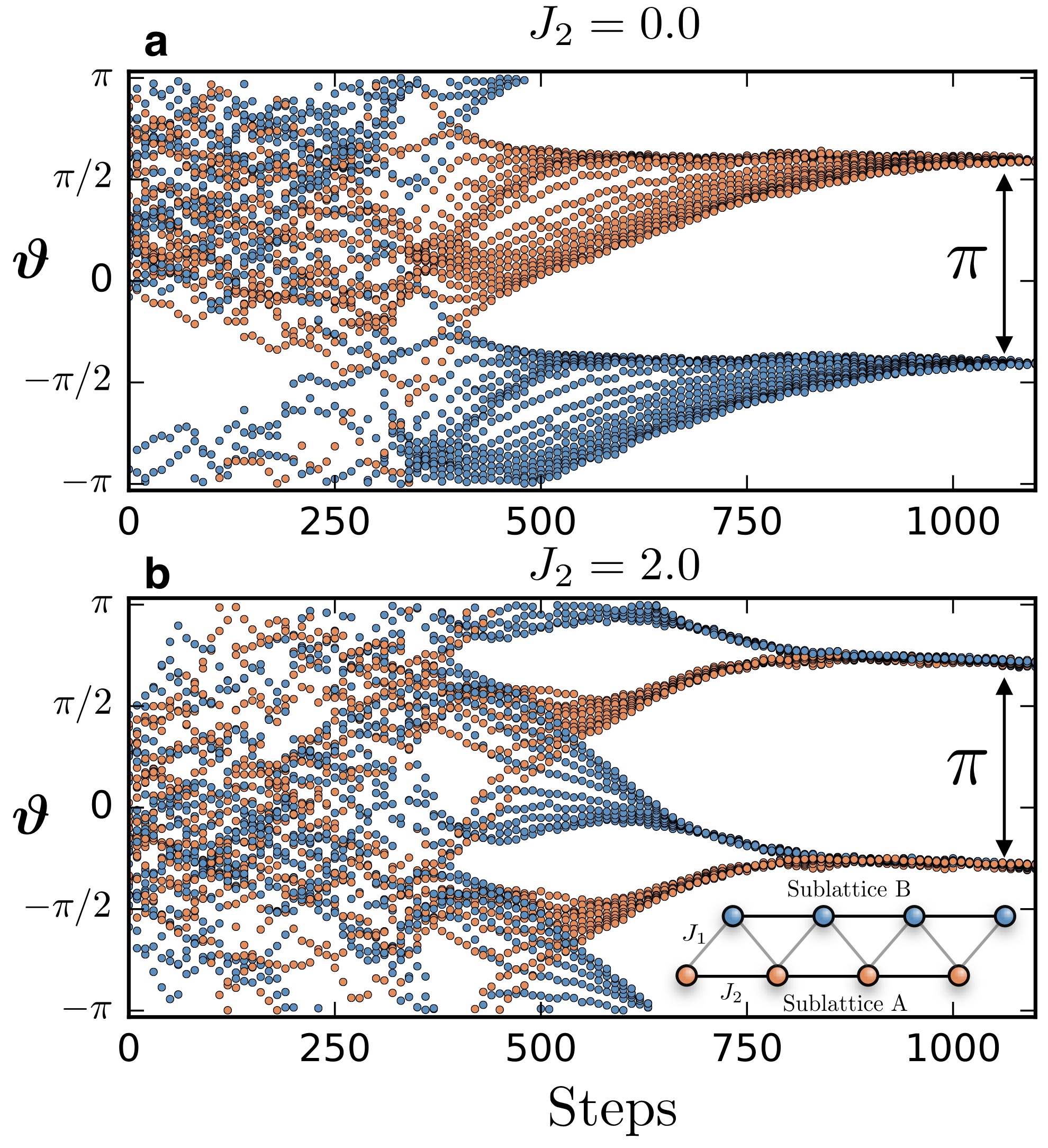}
\caption{Dynamics of the parameters $\bm{\vartheta}$ during training for a ladder of $N=40$ spins ($J_r=0$) with  $J_2=0.0$ ({\bf a}) and $J_2=2.0$ ({\bf b}). The quantum circuit contains only one-qubit rotations around the $z$ axis.}
\label{Fig::j1j2_training} 
\end{figure}

We show the behaviour of the positivization algorithm in Fig.~\ref{Fig::j1j2_training}, where we plot the values of each single rotation angle as a function of the training iteration. For $J_2=0$ (Fig.~\ref{Fig::j1j2_training}a) we observe that all angles corresponding to rotations on sub-lattice $A$ ($B$) converge to the value $\vartheta^{[k]}=\pi/2$ ($-\pi/2$), equivalent to the Marshall sign rule. We then repeat the optimization for an initial ground state obtained by setting $J_2=2.0$ (Fig.~\ref{Fig::j1j2_training}b). Here, we observe that the angles converge to two values separated by $\pi$, but now the rotations on sites from different sublattices are mixed together. It is easy to see that this circuit implements the Marshall sign rule in the limit of $J_1=0$, corresponding to two de-coupled Heisenberg chains. In both cases we measure an average sign of about $0.99$.

Although the relationship is not fully understood, the sign structure of a quantum state is related to its entanglement. For example, a typical random positive wavefunction exhibits a constant law for Renyi entanglement entropies with Renyi index $n>1$, while states with Renyi entropy scaling as a volume law will have a complex sign structure~\cite{Grover05}, suggesting a non-local positivization transformation. 
It therefore stands to reason that circuits of large depths may be required to remove the sign structure when the entanglement needs significant modification. 

In order to increase the entanglement of the starting state, we turn to the exotic spin Bose-Metal (SBM) phase which contains significant entanglement due to the presence of a Bose surface~\cite{Sheng09}. We set $J_2=0$ and examine different initial ground state MPSs obtained for $J_{r}\in[0,1]$, which spans the phase transition into the SBM phase. We optimize circuits with different depths, where a single layer consists of a set of simultaneous commuting two-qubit gates (Fig.~\ref{Fig::qc}). In all simulations, the truncation error in the singular value decompositions performed to restore the MPS representation of the quantum state was kept below $10^{-6}$. 

\begin{figure}[t!]
\noindent \centering{}\includegraphics[width=\columnwidth]{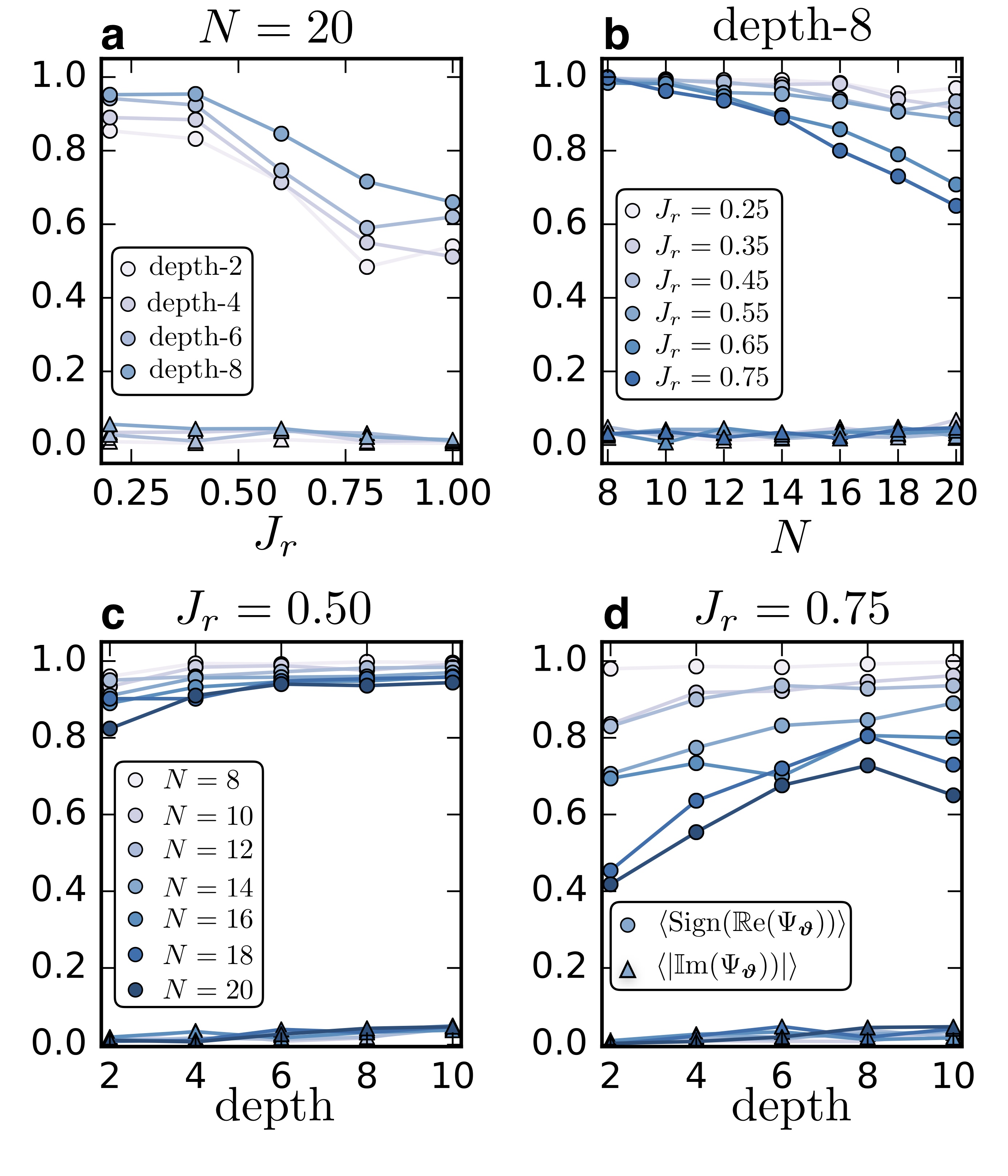}
\caption{Optimization performance for the two-leg triangular Heisenberg chain with ({\bf a}) different depths as a function of $J_r$, ({\bf b}) different values of $J_r$ as a function of the number of spins, and different system sizes as a function of the depth for $J_r=0.25$ ({\bf c}) and $J_r=0.75$ ({\bf d}). The transition into the SBM phase occurs at  $J_{r}\approx0.6$.}
\label{Fig::sbm} 
\end{figure}

We first examine a spin ladder with $N=20$ sites, and optimize circuits of increasing depth for initial ground states obtained at different values of $J_r$. We plot the average sign (circles) and imaginary part (triangles) in Fig.~\ref{Fig::sbm}a. As expected, a larger depth systematically increase the effectiveness of the positivization, which becomes significantly harder as the system is driven into the SBM phase ($J_r\approx0.6$). The transition in complexity is highlighted in Fig.~\ref{Fig::sbm}b, where we show the scalings with the system size for different values of $J_r$ near the critical point for a circuit of fixed depth. In the Bethe phase (small $J_r$), the sign remains sufficiently high as the system size is increased, while the positivization becomes ineffective for larger $N$ in the SBM phase. Finally, we show the scaling against the circuit depth for several sizes $N$ in the two phases of the spin ladder (Fig.~\ref{Fig::sbm}c-d). The results confirm that in the SBM phase, in contrast to the Bethe phase, the depth required to achieve a given average sign increases with the number of spins $N$. In all instances, the optimization succeeds in producing quantum states with  real coefficients to a good approximation.

\paragraph{Conclusions.} 

We have introduced a procedure to systematically search for a local unitary transformation that maps a wavefunction with a sign structure into a non-negative form. The transformation is parametrized as a universal quantum circuit, and the gates are optimized through automatic differentiation algorithms, widely adopted in the machine learning community and implemented with TensorFlow~\cite{tensorflow}. We demonstrated this technique for ground states of a triangular spin ladder with Heisenberg interactions. For the limit of the $J_1$-$J_2$ model, we have shown that the optimization is capable of removing the sign structure, recovering the well-known Marshall sign rule. In the presence of ring-exchange interaction, we observed that the SBM phase demands circuits where the depths scales with the size of the ladders.

The ability to discover a local basis where the average sign of a quantum state becomes substantially higher is particularly relevant for the alleviation of the sign problem in quantum Monte Carlo simulations~\cite{milad2019,Klassen2019,Eisert,gupta2019}. In this context, our positivization algorithm could be repurposed to increase the  ``stoquasticity'' of a target Hamiltonian~\cite{bravyi}. This would require the optimization of a suitably modified cost function, where the input is a matrix product operator representation of the Hamiltonian. This opens interesting prospects for path integrals and projective quantum Monte Carlo simulations, which should be explored in future studies. 

The non-negativity of a wavefunction in a local basis also has direct implications for the data-driven reconstruction of quantum states, which is becoming increasingly important for validating noisy-intermediate scale quantum hardware~\cite{NISQ}. In fact, for wavefunctions with positive amplitudes, experimental data from a single measurement basis is sufficient for the quantum reconstruction of the state, with a particularly favorable scaling with both the system size and the number of measurements~\cite{Torlai18,ml_nisq}.

Finally, our procedure provides a universal, automated method of assessing the complexity associated with the sign problem of a given wavefunction. Ultimately, by performing a systematic finite-size scaling analysis of the resources required to achieve a give average sign, this procedure could be used to determine the complexity class associated with removing the sign structure in various cases, including gapped, critical, or fermionic wavefunctions.  In the future, automated numerical methods based on machine learning technology may be the most promising route to determining the relative ``difficulty'' of various sign structures, and will play a crucial role in formulating a complete theory relating a wavefunction's sign to its 
entanglement structure and simulation complexity.

\subsection*{Acknowledgements }
We thank F. Becca, J. Eisert, M. Ganahl, J. Liu, B. Sanders, M. Stoudenmire and L. Wang for enlightening discussions. The DMRG calculations and the optimization of the circuits were performed using the ITensor~\cite{ITensor} and the TensorFlow~\cite{tensorflow} libraries respectively. This research was supported by the National Science Foundation under Grant No. NSF PHY-1125915.  The Flatiron Institute is supported by the Simons Foundation. M.P.A.F. is grateful to the Heising-Simons Foundation for support. R.G.M. is supported by NSERC, the CRC program, and the Perimeter Institute for Theoretical Physics. Research at Perimeter Institute is supported in part by the Government of Canada through the Department of Innovation, Science and Economic Development Canada and by the Province of Ontario through the Ministry of Economic Development, Job Creation and Trade. J.C. acknowledges support from NSERC and the Canada CIFAR AI chair program.

\bibliographystyle{apsrev4-1}
\bibliography{bibliography.bib}

\newpage
\onecolumngrid
\section*{Supplementary Material}

In this section, we review how to calculate the cost function by exact sampling of a matrix product state (MPS), discuss the main features of automatic differentiation (AD), and show the derivation of the intermediate gradients to efficiently construct the computational graph for optimizing the quantum circuit.

\subsection{The cost function}
The goal of the optimization of the quantum circuit is to discover the set of quantum gate parameters $\bm{\vartheta}$ generating a real and positive final state $\Psi_{\bm{\vartheta}}$. The conditions on $\Psi_{\bm{\vartheta}}$ are:
\begin{equation}
\mathbb{I}\text{m}\big[\Psi_{\bm{\vartheta}}(\bm{\sigma})\big]=0\:\:\:\:\text{and}\:\:\:\: \text{Sign}\big(\mathbb{R}\text{e}\big[\Psi_{\bm{\vartheta}}(\bm{\sigma})\big]\big)=1\:\:\forall\:\:|\bm{\sigma}\rangle\:.
\end{equation}
where $\Psi_{\bm{\vartheta}}(\bm{\sigma})=\langle\bm{\sigma}|\,\hat{\mathcal{U}}_{\bm{\vartheta}}\,|\Psi\rangle$, $\Psi$ is the initial state and $\hat{\mathcal{U}}_{\bm{\vartheta}}$ is the unitary circuit. When translating these conditions into measurable quantities, it is natural to consider their expectation values with respect to the final state $\Psi_{\bm{\vartheta}}$. As clearly neither of them admits a decomposition as a matrix product operator, which would allow calculation by tensor contraction, their expectation values should instead be calculated as averages
\begin{equation}
\sum_{\bm{\sigma}}|\Psi_{\bm{\vartheta}}(\bm{\sigma})|^2\big|\mathbb{I}\text{m}\big[\Psi_{\bm{\vartheta}}(\bm{\sigma})\big]\big|=0\:\:\:\:\text{and}\:\:\:\: \sum_{\bm{\sigma}}|\Psi_{\bm{\vartheta}}(\bm{\sigma})|^2\text{Sign}\big(\mathbb{R}\text{e}\big[\Psi_{\bm{\vartheta}}(\bm{\sigma})\big]\big)=1
\end{equation}
over the probability distribution $p_{\bm{\vartheta}}(\bm{\sigma})=|\Psi_{\bm{\vartheta}}(\bm{\sigma})|^2$. Note that we added the absolute value for the imaginary part in order to avoid a zero average from a distribution centered around zero.

We can define the cost function $\mathcal{C}({\bm{\vartheta}})$ of the optimization as an average over the output distribution
\begin{equation}
\mathcal{C}({\bm{\vartheta}})=\sum_{\bm{\sigma}}|\Psi_{\bm{\vartheta}}(\bm{\sigma})|^2\mathcal{C}_{{\bm{\vartheta}}}(\bm{\sigma})
\label{Eq::cost_generic}
\end{equation}
where $\mathcal{C}_{{\bm{\vartheta}}}(\bm{\sigma})$ combines the two conditions into a convex sum:
\begin{equation}
\begin{split}
\mathcal{C}_{{\bm{\vartheta}}}(\bm{\sigma})=
\gamma\big|\im\big[\Psi_{\bm{\vartheta}}(\bm{\sigma})\big]\big|-(1-\gamma)\text{Sign}\big(\mathbb{R}\text{e}\big[\Psi_{\bm{\vartheta}}(\bm{\sigma})\big]\big)\:.
\end{split}
\end{equation}
with $\gamma\in[0,1]$ (we set $\gamma=0.5$ in our numerical experiments).  Note that the cost function assumes its minimum value $\mathcal{C}_{{\bm{\vartheta}}}(\bm{\sigma}) = \gamma -1$ when the output state is real and its average sign is equal to one. We also point out that the initial average sign can always be made positive by a global transformation. The exponential sum in Eq.~(\ref{Eq::cost_generic}) is then approximated by the average
\begin{equation}
\mathcal{C}({\bm{\vartheta}})\approx \widetilde{\mathcal{C}}(\bm{\vartheta})=\frac{1}{M}\sum_{\bm{\sigma}_j}\mathcal{C}_{{\bm{\vartheta}}}(\bm{\sigma}_j)
\end{equation}
where the configurations $\{\bm{\sigma}_j\}$ are drawn from the probability distribution $p_{\bm{\vartheta}}(\bm{\sigma})$. As a result of the MPS representation of $\Psi_{\bm{\vartheta}}$, it is possible to sample the distribution exactly, leading to a collection of perfectly uncorrelated configurations~\cite{White_mps,Ferris}.

\subsubsection{Perfect sampling of an MPS}
\begin{figure}[t!]
\noindent \centering{}\includegraphics[width=0.8\columnwidth]{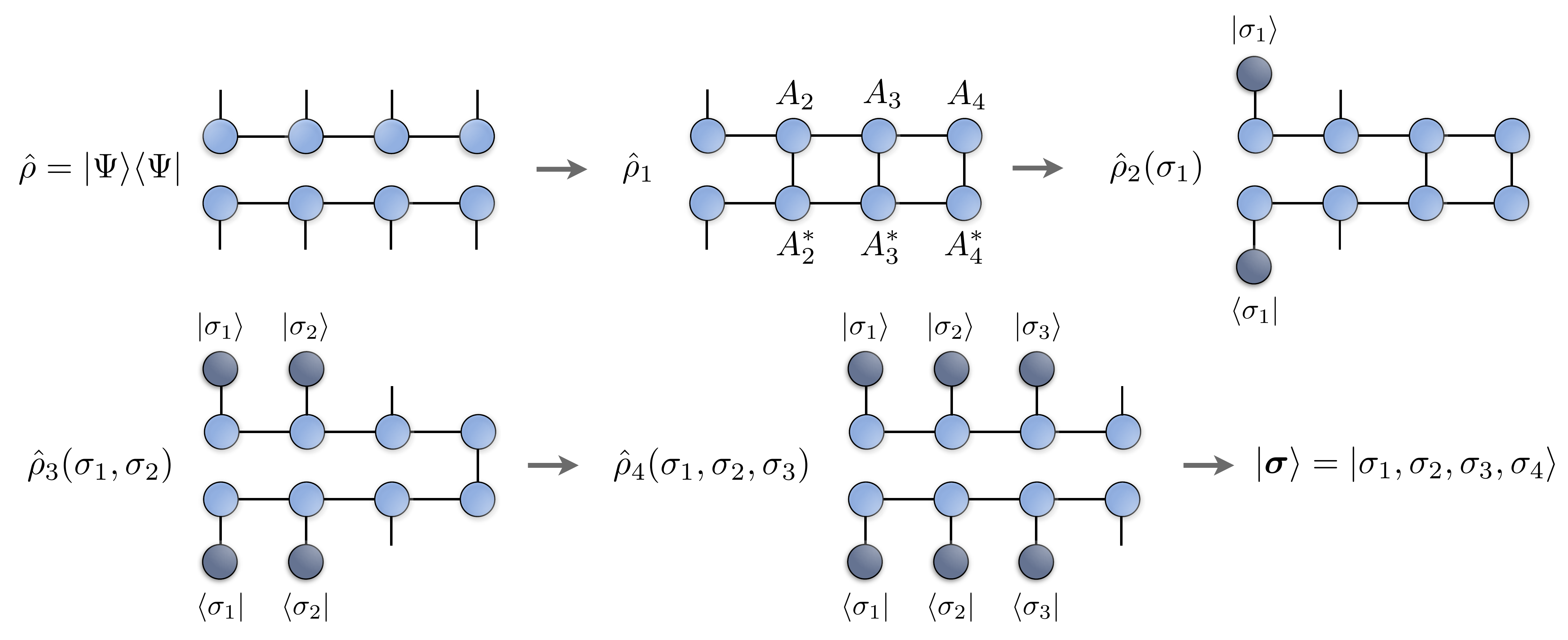}
\caption{Perfect sampling of an MPS with four sites.}
\label{Fig::mps_sampling} 
\end{figure}

The procedure to sample an MPS, schematically pictured in Fig.~\ref{Fig::mps_sampling}, consists of iteratively calculating single-site density matrices, conditional on the state of the sites sampled at the previous step. For the four sites example in Fig.~\ref{Fig::mps_sampling}, one starts from the full density matrix $\hat{\rho}_{\bm{\vartheta}}=|\Psi_{\bm{\vartheta}}\rangle\langle\Psi_{\bm{\vartheta}}|$ and constructs the reduced density matrix for $\sigma_1$:
\begin{equation}
\hat{\rho}_{\bm{\vartheta},1}=\text{Tr}_{\sigma_2,\sigma_3,\sigma_4}|\Psi_{\bm{\vartheta}}\rangle\langle\Psi_{\bm{\vartheta}}|\:.
\end{equation}
The variable $\sigma_1$ is then sampled according to the probability distribution $p_{\bm{\vartheta}}(\sigma_1)=\langle\sigma_1|\hat{\rho}_{\bm{\vartheta},1}|\sigma_1\rangle$, and the density matrix is projected on the subspace corresponding to the measurement outcome:
\begin{equation}
\hat{\rho}_{\bm{\vartheta}}\longrightarrow\hat{\rho}_{\bm{\vartheta}}(\sigma_1)=\frac{\hat{\Pi}_{\sigma_1}\hat{\rho}_{\bm{\vartheta}}\,\hat{\Pi}_{\sigma_1}}{\text{Tr}(\hat{\rho}_{\bm{\vartheta}}\,\hat{\Pi}_{\sigma_1})}\:.
\end{equation}
Next, the single-site density matrix for the state $\sigma_2$, conditional on the state $\sigma_1$, is constructed as
\begin{equation}
\hat{\rho}_{\bm{\vartheta},2}(\sigma_1)=\frac{1}{p_{\bm{\vartheta}}(\sigma_1)}\text{Tr}_{\sigma_3,\sigma_4}\langle\sigma_1|\Psi_{\bm{\vartheta}}\rangle\langle\Psi_{\bm{\vartheta}}|\sigma_1\rangle\:,
\end{equation}
and the state $\sigma_2$ is sampled from the conditional probability distribution $p_{\bm{\vartheta}}(\sigma_2\,|\,\sigma_1)=\langle\sigma_2|\hat{\rho}_{\bm{\vartheta},2}(\sigma_1)|\sigma_2\rangle$, leading to outcome $|\sigma_1,\sigma_2\rangle$ with probability $p_{\bm{\vartheta}}(\sigma_1,\sigma_2)=p_{\bm{\vartheta}}(\sigma_2\,|\,\sigma_1)p_{\bm{\vartheta}}(\sigma_1)$. By repeating this process one obtains a final state $|\bm{\sigma}\rangle$ sampled from the correct probability distribution:
\begin{equation}
p_{\bm{\vartheta}}(\bm{\sigma})=p_{\bm{\vartheta}}(\sigma_N\,|\,\sigma_{N-1},\dots,\sigma_1)\dots p_{\bm{\vartheta}}(\sigma_3\,|\,\sigma_2,\sigma_1)p_{\bm{\vartheta}}(\sigma_2\,|\,\sigma_1)p_{\bm{\vartheta}}(\sigma_1)
=\langle\bm{\sigma}|\hat{\rho}|\bm{\sigma}\rangle=|\langle\bm{\sigma}|\Psi_{\bm{\vartheta}}\rangle|^2\:.
\end{equation}

\subsection{Gradients of the quantum circuit}
Given a specific choice for the structure of the quantum circuit, the optimal values of the parameters $\bm{\vartheta}$ characterizing the quantum gates are found through iterative updates
\begin{equation}
\bm{\vartheta}\longrightarrow\bm{\vartheta}-\eta\,\mathcal{G}(\bm{\vartheta})\:,
\end{equation}
where $\mathcal{G}(\bm{\vartheta})=\nabla_{\bm{\vartheta}}\mathcal{C}(\bm{\vartheta})$ is the gradient of the cost function and $\eta$ is the learning rate (i.e. the step-size of the update). To evaluate the gradient for a generic quantum circuit architecture, we employ the AD framework.

\subsubsection{Automatic differentiation}
The fundamental object in AD is the computational graph, a directed acyclic graph where data (nodes) are processed according to a set of elementary computations (edges)~\cite{AD2000}. Given some input parameters $\bm{\vartheta}$, the output $Y(\bm{\vartheta})$ is obtained through a series of computations $\bm{\vartheta}\rightarrow T_1\rightarrow T_2\rightarrow\dots\rightarrow T_n\rightarrow Y$. The gradient of the output with respect to the input parameters is obtained by applying the chain rule of derivates through the various steps of the computation:
\begin{equation}
\frac{\partial Y}{\partial \bm{\vartheta}}=\frac{\partial Y}{\partial T_n}\frac{\partial T_n}{\partial T_{n-1}}\dots\frac{\partial T_2}{\partial T_1}\frac{\partial T_1}{\partial \bm{\vartheta}}\:.
\end{equation}
Following the common notation in AD, we define $\overline{T}_j=\partial Y / \partial T_j$ as the derivative of the output with respect to the variable $T_j$ in the graph. By considering the particular graph connectivity, this derivative can be expressed in terms of the derivative of adjacent variables in the graph:
\begin{equation}
\overline{T}_j=\sum_{i\in\mathscr{N}(j)}\overline{T}_i\frac{\partial T_i}{\partial T_j}\:,
\end{equation}
where $\mathscr{N}(j)$ is the set of nodes identifying the variables connected to $T_j$. All derivatives can then be evaluated by performing a forward pass in the graph (storing the various results of the computations), followed by a backward pass where derivatives with respect to each node are calculated based on a pre-determined set of primitives, such as simple functions and linear algebra routines. This type of AD algorithm is called {\it reverse-accumulation}, since it starts the calculation of the gradients from the output variable $Y$. It is also possible to calculate the gradients using a forward mode, i.e. propagating derivates from the input to the output. This AD mode is however less efficient when the input dimension is much larger than the output dimension. Since that is often the case, reverse-accumulation AD is most widely used in practical applications.

\subsubsection{Intermediate gradients}

The AD framework can be readily applied to optimize tensor networks and  quantum circuits~\cite{Liao19}, but care should be taken in our specific case. In fact, instead of calculating the cost function by tensor contraction, we express it as the average
\begin{equation}
\mathcal{C}({\bm{\vartheta}})=\sum_{\bm{\sigma}}|\Psi_{\bm{\vartheta}}(\bm{\sigma})|^2\mathcal{C}_{{\bm{\vartheta}}}(\bm{\sigma})
\end{equation}
over the probability distribution $p_{\bm{\vartheta}}(\bm{\sigma})=|\Psi_{\bm{\vartheta}}(\bm{\sigma})|^2$. As such, AD of $\mathcal{C}({\bm{\vartheta}})$ would quickly become intractable given the exponential size of the sum in the number of sites $N$. In contrast, since we are able to sample from $p_{\bm{\vartheta}}(\bm{\sigma})$, we can approximate the cost function with the finite-size average
\begin{equation}
\mathcal{C}({\bm{\vartheta}})\approx \frac{1}{M}\sum_{\bm{\sigma}_j}\mathcal{C}_{{\bm{\vartheta}}}(\bm{\sigma}_j)\equiv\widetilde{\mathcal{C}}(\bm{\vartheta})\:.
\end{equation}
However, we note that by implementing a computational graph $\bm{\vartheta}\rightarrow\widetilde{\mathcal{C}}(\bm{\vartheta})$, that is by approximating the distribution $p_{\bm{\vartheta}}(\bm{\sigma})$ with the samples, we are neglecting its parametric dependence on $\bm{\vartheta}$, which has a non-trivial contribution to the cost function and its gradients. Specifically, the gradients obtained by applying AD to the graph do not correspond to the actual gradients of the cost function
\begin{equation}
\nabla_{\bm{\vartheta}}\widetilde{\mathcal{C}}(\bm{\vartheta})=\frac{1}{M}\sum_{\bm{\sigma}_j}\nabla_{\bm{\vartheta}}\mathcal{C}_{{\bm{\vartheta}}}(\bm{\sigma}_j)\ne\mathcal{G}(\bm{\vartheta})=\nabla_{\bm{\vartheta}}\mathcal{C}(\bm{\vartheta})\:.
\label{Eq::cost_mc}
\end{equation}
The inequality does not follow from statistical uncertainty deriving from averaging over samples, but rather from the elimination of the dependence of the underlying distribution from $\widetilde{\mathcal{C}}(\bm{\vartheta})$. 

To solve this issue, the gradients need to be calculated analytically until the dependence on the distribution is eliminated. At that point, AD can be safely applied. We start by taking the gradient of the full cost function, without any approximation:
\begin{equation}
\begin{split}
\mathcal{G}(\bm{\vartheta})&=\nabla_{\bm{\vartheta}}\mathcal{C}(\bm{\vartheta})=\nabla_{\bm{\vartheta}}\sum_{\bm{\sigma}}|\Psi_{\bm{\vartheta}}(\bm{\sigma})|^2\mathcal{C}_{{\bm{\vartheta}}}(\bm{\sigma})=
\sum_{\bm{\sigma}}\Big[|\Psi_{\bm{\vartheta}}(\bm{\sigma})|^2\nabla_{\bm{\vartheta}}\mathcal{C}_{{\bm{\vartheta}}}(\bm{\sigma})+
\mathcal{C}_{{\bm{\vartheta}}}(\bm{\sigma})\nabla_{\bm{\vartheta}}\big(\Psi_{\bm{\vartheta}}(\bm{\sigma})\Psi^*_{\bm{\vartheta}}(\bm{\sigma})\big)\Big]\\
&=\sum_{\bm{\sigma}}\Big[|\Psi_{\bm{\vartheta}}(\bm{\sigma})|^2\nabla_{\bm{\vartheta}}\mathcal{C}_{{\bm{\vartheta}}}(\bm{\sigma})+
\mathcal{C}_{{\bm{\vartheta}}}(\bm{\sigma})\big(\Psi^*_{\bm{\vartheta}}(\bm{\sigma})\nabla_{\bm{\vartheta}}\Psi_{\bm{\vartheta}}(\bm{\sigma})+\Psi_{\bm{\vartheta}}(\bm{\sigma})\nabla_{\bm{\ vartheta}}\Psi^*_{\bm{\vartheta}}(\bm{\sigma})\big)\Big]\\
&=\sum_{\bm{\sigma}}|\Psi_{\bm{\vartheta}}(\bm{\sigma})|^2\Big[\nabla_{\bm{\vartheta}}\mathcal{C}_{{\bm{\vartheta}}}(\bm{\sigma})+
2\,\mathcal{C}_{{\bm{\vartheta}}}(\bm{\sigma})\re\big[\nabla_{\bm{\vartheta}}\log\Psi_{\bm{\vartheta}}(\bm{\sigma})\big]\Big]\:.
\end{split}
\end{equation}
Since the gradient $\mathcal{G}(\bm{\vartheta})$ is still in the form of an average with respect to the probability distribution $p_{\bm{\vartheta}}(\bm{\sigma})$, we can now approximate the above equation with a sum over a finite number of samples:
\begin{equation}
\mathcal{G}(\bm{\vartheta})\approx\widetilde{\mathcal{G}}(\bm{\vartheta})=
\frac{1}{M}\sum_{\bm{\sigma}_j}\Big[\nabla_{\bm{\vartheta}}\mathcal{C}_{{\bm{\vartheta}}}(\bm{\sigma}_j)+
2\,\mathcal{C}_{{\bm{\vartheta}}}(\bm{\sigma}_j)\re\big[\nabla_{\bm{\vartheta}}\log\Psi_{\bm{\vartheta}}(\bm{\sigma}_j)\big]\Big]\:.
\label{Eq::grad_ex}
\end{equation}
Note that $\widetilde{\mathcal{G}}(\bm{\vartheta})\ne\nabla_{\bm{\vartheta}}\widetilde{\mathcal{C}}(\bm{\vartheta})$ since the approximation of the average with a finite number of samples has now been taken after applying the gradient operation. We can then simply write
\begin{equation}
\widetilde{\mathcal{G}}(\bm{\vartheta})=\frac{1}{M}\sum_{\bm{\sigma}_j}\mathcal{G}_{\bm{\vartheta}}(\bm{\sigma}_j)
\end{equation}
with $\mathcal{G}_{\bm{\vartheta}}(\bm{\sigma}_j)=\nabla_{\bm{\vartheta}}\mathcal{C}^\star_{\bm{\vartheta}}(\bm{\sigma}_j)$, for a new {\it effective} sample-wise cost function
\begin{equation}
\mathcal{C}^\star_{\bm{\vartheta}}(\bm{\sigma}_j)=\mathcal{C}_{\bm{\vartheta}}(\bm{\sigma})+2\{\mathcal{C}_{\bm{\vartheta}}(\bm{\sigma})\}_{\text{\footnotesize{ng}}}\:\re\big[\log\Psi_{\bm{\vartheta}}(\bm{\sigma})\big]\:,
\end{equation}
where $\{\cdot\}_{\text{\footnotesize{ng}}}$ means that the argument should not be differentiated. It is straightforward to see that, by applying AD to the cost function
\begin{equation}
\widetilde{\mathcal{C}}^\star(\bm{\vartheta})=\frac{1}{M}\sum_{\bm{\sigma}_j}\mathcal{C}^\star_{\bm{\vartheta}}(\bm{\sigma}_j)\:,
\end{equation}
we obtain the correct approximation of the gradients $\widetilde{\mathcal{G}}(\bm{\vartheta})\approx\mathcal{G}(\bm{\vartheta})$ (within statistical uncertainty). Finally, to make the cost function differentiable, we replace the sign function with a ``soft sign'' $Sign(x)=2S_{\beta}(x)-1$, where
\begin{equation}
S_{\beta}(x)=\frac{1}{1+e^{-\beta x}}
\end{equation}
and the fictitious inverse temperature $\beta$ controls the softness of the sign.

\end{document}